\documentclass[sigconf,screen]{acmart}

\renewcommand\footnotetextcopyrightpermission[1]{}

\AtBeginDocument{%
  \providecommand\BibTeX{{%
    \normalfont B\kern-0.5em{\scshape i\kern-0.25em b}\kern-0.8em\TeX}}}

\usepackage[utf8]{inputenc} 
\usepackage[T1]{fontenc}    
\usepackage{hyperref}       
\usepackage{url}            
\usepackage{booktabs}       
\usepackage{amsfonts}       
\usepackage{nicefrac}       
\usepackage{microtype}      
\usepackage{graphicx}
\usepackage{amsmath,amsfonts,amsthm}
\usepackage{array}
\usepackage{subcaption}
\usepackage{bbm}
\usepackage{bm}
\usepackage{multicol}
\usepackage{lipsum}
\usepackage{wrapfig}
\usepackage[ruled]{algorithm2e}
\usepackage{arydshln}
\usepackage{multirow}
\usepackage{enumitem}
\usepackage{siunitx}
\usepackage{dsfont}

\newtheorem{lemma}{Lemma}

\newcommand{\Ical}{\mathcal{I}}

\newcommand{\Hcal}{\mathcal{H}}

\newcommand{\Ocal}{\mathcal{O}}
\newcommand{\Acal}{\mathcal{A}}

\newcommand{\Ebb}{\mathbb{E}}

\newcommand{\Dbb}{\mathbb{D}}

\newcommand{\hbf}{\mathbf{h}}

\begin{document}

\title{On the Advances and Challenges of Adaptive Online Testing}


\author{Da Xu$^*$}
\affiliation{%
  \institution{Walmart Labs}
  \city{Sunnyvale}
  \state{California}
  \country{USA}
}
\email{daxu5180@gmail.com}

\author{Bo Yang}
\authornote{The two authors contribute equally to this work.}
\affiliation{%
  \institution{LinkedIn Corporation}
  \city{Sunnyvale}
  \state{California}
  \country{USA}
}
\email{by9ru@virginia.edu}

\copyrightyear{2022}
\acmYear{2022}
\setcopyright{acmlicensed}\acmConference[WSDM '22]{Workshop on Decision Making for Modern Information Retrieval System}{February 21--25, 2022}{Tempe, AZ, USA}
\acmBooktitle{Workshop on Decision Making for Modern Information Retrieval System (WSDM '22), February 21--25, 2022, Tempe, AZ, USA}

\begin{abstract}
In recent years, the interest in developing adaptive solutions for online testing has grown significantly in the industry.
While the advances related to this relative new technology have been developed in multiple domains, it lacks in the literature a systematic and complete treatment of the procedure that involves exploration, inference, and analysis. 
This short paper aims to develop a comprehensive understanding of adaptive online testing, including various building blocks and analytical results. We also address the latest developments, research directions, and challenges that have been less mentioned in the literature. 
\end{abstract}



\maketitle

\section{Introduction}
\label{sec:intro}
Decision making in modern industry is often driven by randomized experiments behind the scenes. In particular for IT companies, the production scenario ranges from deciding the best-performing recommendation system to determining the size of an ad banner. The traditional process of simultaneously launching alternative services to randomly assigned users is also known as \emph{online A/B/n testing} \cite{kohavi2017online,fabijan2018online}.
While A/B/n testing and its variants enjoy a high degree of simplicity in planning and making inference, they still lack adaptability when certain alternatives become clearly inferior. Perhaps the most critical consequence would lost revenue and user satisfaction, so optimizing online testing strategy has become imperative for many, if not all.

We start our discussion by first characterizing an online experimentation. By convention, we use the notion of \emph{arm} to indicate such as a candidate webpage service, denote by: $a \in \Acal$ with $|\Acal|=k+1$. Sampling (obtaining) an observation from arm $a$ is represented by: $X_{a,i} \sim P_a, i\in\{1,\ldots,n_a\}$, where $P_a$ is the reward (e.g. impression rate, revenue, return of investment) distribution associated with arm $a$. We use $a_0$ denote the \emph{control arm} which often refers to the currently deployed service. We will follow this convention unless specified. While $X_{a,i}$ are often assumed to be identically distributed under each arm $a$, they are not necessarily independent. Handling dependency will be an important topic in our following discussion, since it cannot be avoided by adaptive exploration, but would violate traditional statistics inference assumptions \cite{casella2021statistical}. 

For the most part, the inference goal of an online experiment is to test a statistical hypothesis often related to the data distribution, e.g. whether or not $\mu_{1} > \mu_2$ where $\mu_q := \Ebb[X_{a_q,i}]$ is the expected reward for arm $q$. The outcome of a hypothesis testing is usually expressed via the null hypothesis $H_0$, alternative hypothesis $H_1$, a significance level $\alpha \in (0,1)$, and the observed samples. For example, when $H_0$ is rejected, it holds: 
\begin{equation}
\label{eqn:hypothesis-testing}
Pr\big(\text{rejecting } H_0 \text{ for } H_1 \,|\, H_0 \text{ is true}\big) \leq \alpha,
\end{equation}
In other words, it is unlikely that $H_1$ better characterizes the observed data distribution just by chance. Also, the control arm $a_0$ usually plays a pivotal role in the hypothesis's design: $a_0$ is often assumed to be superior unless overturned by enough evidence.

Practically for IT companies, adaptive online testing means the proportion of traffic guided to each arm during the random assignment can be adjusted based on the performance to date. Since the purpose of online testing is purely \emph{exploratory}, i.e. to explore as much as possible without worrying about the cost, the problem setting differs naturally from \emph{bandit} or \emph{reinforcement learning} where exploitation is mandatory \cite{sutton2018reinforcement,auer2002finite}. Nevertheless, some exploration strategies from the bandit literature also suit the pure exploration task \cite{lattimore2020bandit}, and they are often studied under the \emph{best-arm identification} problem \cite{audibert2010best,paulson1964sequential}. 

In particular, suppose $a^{[i]}$ is the arm with the $i^{\text{th}}$ highest reward, and denote by: $\Delta_j=\mu_{a^{[1]}} - \mu_{a_j}$ which is the \emph{suboptimality gap} of arm $a_j$. We use $\hbf_{T}$ to represent the first $T$ samples where $\hbf_{T}\in\Hcal_{T}$. The exploratory algorithms from best-arm identification can be described by a stopping time\footnote{We assume the measurability of $T$ is granted throughout our discussion.} $T$ and a data-adaptive decision rule $\Dbb: \Hcal_{T}\to [k]$ such that when the algorithm halts, the probability that $a^{[1]}$ is not identified (which equivalents to the event of $\Delta_{\Dbb(\hbf_{T})} > 0$) becomes small:
\begin{equation}
\label{eqn:best-arm-id}
Pr\big(T < \infty \text{ and } \Delta_{\Dbb(\hbf_{T})} > 0\big) \leq \delta,
\end{equation}
where $\delta \in (0,1)$ is also some pre-defined confidence level. 

While best-arm identification and hypothesis testing are similar in finding the best candidate (hypothesis or arm) with some tolerance of making mistake due to randomness, their meanings are ultimately different. One apparent reason is that the probability in (\ref{eqn:best-arm-id}) is taken w.r.t. the sampling and reward distributions, while the conditional probability in (\ref{eqn:hypothesis-testing}) is taken w.r.t. the distribution under $H_0$. The conclusion of one problem does not directly imply another.
Meanwhile, notice that as long as the correct statistical procedure is employed, the hypothesis testing does not necessarily depend on the sampling strategy. 
There is possibility to employ the sampling strategy from best-arm identification and use hypothesis testing as part of the stopping rule. This integration may bring adaptability to online testing while not costing its statistical rigorousness.
Now both problems will be subject to new changes:
\begin{itemize}[leftmargin=*]
    \item the current sample may depend on the past ones due to the adaptive sampling strategy;
    \item there will be continuous interactions between the stopping criteria (and stopping time) of the best-arm identification, the p-value, and the significance level of hypothesis testing.
\end{itemize}
The first change rules out traditional hypothesis testing procedures (such as the \emph{Student's t-test}) since the observations are no longer i.i.d. The impact of the second change is more involved since it concerns the working mechanisms of both best-arm identification and hypothesis testing. The good news is we can leverage \emph{confidence bounds} to connect the decision-makings of best-arm identification with hypothesis testing.
Nevertheless, the nature of p-value and significance level will be different in the sequential setting, and they need to be adjusted along the path of adaptive exploration. 
Finally, even the best strategy may struggle to find the best-performing arm if the structure of a particular problem instance is unfriendly. Studying this \emph{instance-wise} difficulty would be valuable for understanding the inherent challenges of adaptive online testing.  


Section \ref{sec:background} covers the basics of pure exploration and hypothesis testing with a highlight on confidence interval. We study in Section \ref{sec:advances} the more advanced tools and solutions for integrating interact hypothesis testing with best-arm identification. In Section \ref{sec:difficulty}, we discuss the difficulty of best-arm identification and the challenges for adaptive online testing. We may omit technical details for the sake of space, and readers may refer to the references provided.

\textbf{Contribution:} since our paper is the first in IR literature that systematically discuss exploration, inference, and analysis of adaptive online testing, it aims to serve both as introductory material for unfamiliar readers and a technical summary for domain experts.


\section{Pure exploration, confidence interval, and p-value}
\label{sec:background}
The best-arm identification in (\ref{eqn:best-arm-id}), which we described as a pure exploration problem, takes the \emph{fixed confidence} setup\footnote{There are two related but different settings in best-arm identification: fixed confidence and fixed budget. The solutions can often be switched between the two settings \cite{gabillon2012best}, so we stick with the fixed confidence problem which has been studied in more depth.} \cite{garivier2016optimal}.
Using exploration strategy to optimize online problems means finding the optimal arm with as few samples as possible. 
In particular, the strategy must be carefully designed so each arm is explored just enough to be separated from the others with high confidence. 
The traditional A/B/n testing acts like a \emph{uniform exploration} policy where the stopping rule is to reach the maximum round of experiments or reject the null hypothesis. 
Probably the most important idea behind many pure exploration algorithms is the \emph{confidence bound} (CB) method \cite{auer2002finite}, where empirical uncertainty is used to construct optimistic/pessimistic overestimations of the mean reward.

Note that the randomness in finite samples can cause any estimation of $\big\{\mu_i\big\}_{i=1}^k$ subject to uncertainty. 
Under mild distribution assumptions such as $P_a$ is \emph{Gaussian} or \emph{sub-Gaussian}, there exists a (provably tight) region that contains $\mu_i$ with arbitrarily high probability. 
The region is often constructed using finite-sample concentration results such as the \emph{Hoeffding} or \emph{Chernoff} inequalities \cite{boucheron2013concentration}, which we exemplify as below.
\begin{lemma}[Concentration of i.i.d. mean]
\label{lemma:hoeffding}
Assume $X_{a,i}, a=0,\ldots,k$ are independent, $\sigma$-sub-Gaussian random variables. Then for any $\epsilon>0$:
\begin{equation}
\label{eqn:hoeffding}
    Pr\Big(\frac{1}{n_a}\sum_{i=1}^{n_a} X_{a,i} - \mu_a \geq \epsilon \Big) \leq  \exp(\frac{-n_a\epsilon^2}{2\sigma^2}),
\end{equation}
and the same bound holds for $Pr(\frac{1}{n_a} \sum_{i=1}^{n_a} X_{a,i}-\mu_a \leq -\epsilon)$.
\end{lemma}

We immediately observe that by solving for $\exp(\frac{-n_a\epsilon^2}{2\sigma^2})=\frac{\delta}{2}$, it holds with probability at least $1-\delta$:
\begin{equation}
\mu_a \in \big[\underbrace{\hat{\mu}_a - \sqrt{2\sigma^2\log(1/\delta)/n_a}}_{\text{LCB}^{(a)}(n_a,\delta)} \,,\, \underbrace{\hat{\mu}_a + \sqrt{2\sigma^2\log(1/\delta)/n_a}}_{\text{UCB}^{(a)}(n_a,\delta)}\big],
\end{equation}
where $\hat{\mu}_a = 1/n_a \cdot \sum_{i=1}^{n_a} x_{a,i}$, and \emph{LCB} and \emph{UBC} are shorthands for the lower/upper confidence bound. 
Both $UCB$ and $LCB$ express as: $\hat{\mu} \pm \text{CI}(n_a,\delta)$, where $CI$ is the confidence interval that depends on the sample size and confidence level $\delta$. 

For any two arms $a_1$ and $a_2$, $\text{LCB}^{(a_1)}$ and $\text{UCB}^{(a_2)}$ provide the pessimistic and optimistic estimation of $\mu_1$ and $\mu_2$, respectively. If $\text{LCB}^{(a_1)} > \text{UCB}^{(a_2)}$, it is easy to show that $\mu_1>\mu_2$ with high probability, again by using concentration inequalities. 
Therefore, the CB method provides an elegant but powerful approach to tailor decision making based on the uncertainty of an estimation. 
The two major algorithms for best-arm identification, namely \emph{Action Elimination} and \emph{Upper Confidence Bound}, both rely on CB:
\begin{itemize}[leftmargin=*]
    \item \textbf{Action Elimination} \cite{even2002pac}: sample from each arm a pre-determined $n_{\text{init}}$ times, and eliminate according to: $\text{UCB}^{(a_i)} < \text{LCB}^{(a_0)}$ where $a_0$ is a reference arm. Continue until only one arm is left.
    \item \textbf{Upper Confidence Bound} \cite{audibert2010best,jamieson2014lil}: keep sampling from the arm indexed by $\arg\max_a \text{UCB}^{(a)}$, and stop when there exists an $a^*$ such that: $\text{LCB}^{(a^*)} \geq \text{UCB}^{(a_i)}$ for all $i=1,\ldots,k$.
\end{itemize}

We point out that both approaches have their limitations when used as an exploration strategy for online testing.
The merit of Action Elimination is to throw away arms only after we accumulate enough confidence on their suboptimality. 
Although it improves upon the uniform exploration, the suboptimal arms is still likely to be explored more than necessary. 
The Upper Confidence Bound is on the opposite side: the leading arm will be overexplored, and even the second-best arm may be underexplored. Luckily, both of them can be improved by using an variant of the Upper Confidence Bound method, which is known as \emph{LUCB}. LUCB aims to find a better balance between exploring the leading arm and others:
\begin{itemize}[leftmargin=*]
    \item \textbf{LUCB} \cite{kaufmann2013information,kalyanakrishnan2012pac}: keep sampling from both the current \emph{best} and \emph{second-best} arm (denote by $a^*$ and $a^{**}$) in terms of $\big\{UCB^{(i)}\big\}_{i=1}^k$,  and stop when: $\text{LCB}^{(a^*)} > \text{UCB}^{(a^{**})}$.
\end{itemize}

For all the above algorithms, confidence intervals not only guides how to explore, but also decides when to stop. 
As we show next, this will be an advantage for bringing in hypothesis testing, because confidence intervals can also be related to p-value.

Although criticized for biasing scientists and researchers toward a particular perspective of experimentation \cite{amrhein2019scientists}, \emph{p-value} remains an important metric to understand the significance of a discovery. Generally speaking, p-value is the probability of obtaining test results at least as extreme as what is observed, if the null hypothesis $H_0$ is correct. It usually involves a \emph{test statistic} $Z$ whose distribution is known under $H_0$. Suppose the observed test statistic is $z$, then the probability of having more extreme value on both sides of the distribution is given by: $Pr\big(|Z>z| \,\big|\,H_0\big)$.

When the quantity of interest is mean reward, one straightforward choice is letting: $Z_{k}:=\frac{1}{n_{0}}\sum_{i=1}^{n_{0}}X_{a_0,i} - \frac{1}{n_{k}}\sum_{i=1}^{n_{k}}X_{a_k,i} $. Under the null hypothesis $H_0: \mu_{a_0} = \mu_{a_k}$, for independent (sub)-Gaussian $X_i$, the test statistic $Z_k$ follows a \emph{zero-mean} (sub)-Gaussian distribution. If the distribution of $Z_k$ is symmetric, the p-value is:
\[
Pr\big(|Z_k>z_k| \,\big|\, H_0\big) = Pr\big(\Big|\frac{1}{n_{0}}\sum_{i=1}^{n_{0}}X_{a_0,i} - \frac{1}{n_{k}}\sum_{i=1}^{n_{k}}X_{a_k,i}\Big| > |z_k| \big).
\]
As an immediate result, p-values can also be obtained from concentration inequalities similar to the ones in Lemma \ref{lemma:hoeffding}, since they both measure the deviations under some fixed distribution.
Following the same logic as CB, it is easy to check that $Pr\big(Z_k>z_k \,\big|\, H_0\big) \leq \alpha$ (which $\alpha$ is the pre-defined significance level) also leads to a confidence-interval-based decision making, e.g.:
\[
\text{reject }H_0 \text{ if } z_k \not\in \big[{-\text{CB}}(n_k,n_0,\alpha), {\text{CB}}(n_k,n_0,\alpha)\big],
\]
where ${\text{CB}}$ is the associated confidence interval whose expression we omit here for brevity. In fact, interested readers may check that the single-arm p-values for comparing $a_0$ and $a_k$ under $H_0: \mu_{a_0} \geq \mu_{a_k}$ can be obtained by inverting $LCB$ and $UCB$ via \cite{keener2010theoretical}:
\begin{equation}
\label{eqn:p-value}
P_{n_a,n_k} = \sup \big\{\gamma\in[0,1] \,\big|\, \text{LCB}^{(a_k)}(n_k,\gamma) \geq \text{UCB}^{(a_0)}(n_0,\gamma) \big\}.
\end{equation}

The above link between p-value and confidence interval demonstrate the feasibility of combining best-arm identification with hypothesis testing. In particular, there could exist a unified decision-making procedure using only confidence intervals.
However, there are several obstacles that require further investigation, notably:
\begin{enumerate}[leftmargin=*]
    \item how to obtain valid confidence intervals when the observations are no longer independent;
    \item how to make sure the p-value is always valid and the false discovery rate is controlled, since the testing can be queried by the adaptive exploration algorithm at any time.
\end{enumerate}
In the next section, we discuss some of the more advanced tools and methodologies that may address the above challenges.

\section{Advances in best-arm identification and hypothesis testing}
\label{sec:advances}
When $X_{i}$ are not independent, or when the sample size $n(T)$ itself is a random variable that depends on some stopping time, $\sum_{i=1}^{n}X_{a,i}$ becomes a \emph{random walk} whose concentration behavior differs from the Hoeffding-type inequalities we introduced earlier. 
Historically, the \emph{law of iterated logarithm} (LIL) \cite{hartman1941law} was developed to describe the asymptotic behavior of a zero-mean unit-variance random walk:
\begin{equation}
\label{eqn:LIL}
\underset{n\to\infty}{\lim\sup} \frac{\sum_{i=1}^n X_i}{\sqrt{2n\log\log n}} = 1,\quad \text{almost surely}.
\end{equation}
Unfortunately, LIL only characterizes the limiting behavior, while real-world online testings often falls into the finite-sample regime. The recent work by \citet{zhao2016adaptive} propose a novel finite-sample bound for such cases, from which we can construct confidence intervals for both best-arm identification and hypothesis testing.
\begin{lemma}[Concentration of random walk]
\label{lemma:adaptive-hoeffding}
Let $T$ be any random variable taking value in $\mathbb{N}$. Let $S_{T}:=\sum_{i=1}^{T}X_i$. There exists $a>0$ such that:
\begin{equation}
\label{eqn:adaptive-hoeffding}
Pr\Big(S_{T} \geq \sqrt{0.6n\log(\log_{1.1}n+1) + an} \Big) \leq 12\exp(-1.8a).
\end{equation}
\end{lemma}

Comparing the convergence rates implied by Lemma \ref{lemma:hoeffding} and \ref{lemma:adaptive-hoeffding}, we spot a $\sqrt{\log\log n}$ difference caused by the extra stochasticity from dependent observations. 
This also holds asymptotically if we compare LIL with the \emph{law of large numbers}. As a result, the confidence intervals for adaptive online testing may just be wider than its non-adaptive counterpart. It means that under the same context, more observations may be required to trigger a decision.

Now that we know how to compute confidence bounds correctly, the next challenge is to make p-values \emph{valid at all times}. 
This step is critical if the stopping rule of online exploration can consult the p-value at any time. 
Just like having a monkey randomly tap a typewriter indefinitely, it will almost surely enter any given text. 
Ideally, an any-time p-value should allow guilt-free queries from the continuing adaptive exploration strategy, and the previous study by \citet{johari2015always} suggests that a always valid p-value is a stochastic process $\{P^{(t)}\}_{t=1}^{\infty}$ such that for any (random) stopping time $T$, under any distribution $\mathbb{P}_{\Hcal_0}$ that agrees with the null hypothesis, we have:
\begin{equation}
\label{eqn:super-unif}
    \mathbb{P}_{\Hcal_0}(P^{(T)}\leq \alpha) \leq \alpha.
\end{equation}
Note that this definition extends a classical property of p-value that it should follows \emph{uniform distribution} under $\mathbb{P}_{\Hcal_0}$.

Recently, \citet{yang2017framework} proposes taking the \emph{minimum} of all single-arm p-value, from all the previous steps, to obtain an \emph{any-time} p-value such that it satisfies (\ref{eqn:super-unif}). 
For instance, at step $T$, let $n_0(T)$ and $n_k(T)$ be the $\#$samples from arm $a_0$ and $a_k$ collected until the $T^{\text{th}}$ step, then the any-time p-value for the experiment is:
\[
P^{(T)}_{a_0,a_k} := \min_{s\leq T}\min_{i=1,\ldots,k}P_{n_0(T),n_i(T)}.
\]
The idea behind taking all the minimums is to make sure that:
$
\mathbb{P}\big(\bigcup_{j=1}^k\bigcup_{t=1}^{\infty}\big\{P_{n_0(t),n_j(t)}\leq \alpha\big\}\big)\leq\alpha
$, so $P^{(T)}$ is always valid.

It is also known that controlling the false alarm error of each test is not sufficient to achieve a small \emph{false discovery rate} (FDR) when multiple testings occur. Let $\alpha^{(t)}$ be the significance level at step $t$. When running an adaptive online testing, $1\big[P^{(t)} < \alpha^{(t)}\big]$ will be queried multiple times. Therefore, to control FDR in an online fashion (we use $R^{(t)}=1$ to denote claiming a discovery at step $t$), it is intuitive that $\alpha^{(t)}$ should be aware of the history $\{R^{(j)}\}_{j=1}^{(t-1)}$: 
\[
\alpha^{(t)} := \alpha^{(t)}\big(\alpha;R^{(1)},\ldots,R^{(t-1)}\big),
\]
by using a proper decision rule $\alpha^{(t)}$. Formally, the FDR at step $t$ is defined as $\Ebb\Big[\frac{\#\text{True discoveries}}{\#\text{Discoveries made before }t} \Big]$. In the non-adaptive setting, the \emph{Benjamini-Hochberg} procedure is a standard way to control FDR \cite{benjamini1995controlling}. 
For adaptive experimentation, we find the idea of \emph{"$\alpha$-investing"} powerful for controlling online FDR \cite{foster2008alpha}. The general idea is to let the decision rule $\alpha^{(t)}$ "invest" a small amount of confidence each time when a testing is called, and $\alpha^{(t)}$ will gain some reward if a discovery is claimed. By choosing the proper investment and reward function, this procedure has been shown to achieve FDR control. The implementation is quite straightforward and readers may refer to the references herein \cite{javanmard2018online,aharoni2014generalized,ramdas2017online}. 

We now put together the pieces and present an exemplary adaptive online testing procedure. We first sample from each arm $n_{\text{init}}$ times for warm-start. After that, at each step $t$: 

\begin{enumerate}[leftmargin=*]
    \item Obtain the adjusted significance level $\alpha^{(t)}$ using the history p-values $\{P^{(j)}\}_{j=1}^{t-1}$ and rejections $\{R^{(j)}\}_{j=1}^{t-1}$, with $\alpha$-investing.
    \item First follow LCUB: compute $\big\{\text{UCB}^{a_i}\big(n_i(t),\alpha^{(t)}\big)\big\}_{i=0}^k$, find $a^*$ and $a^{**}$. If not $\text{LCB}^{a^*}>\text{UCB}^{a^{**}}$ and $\text{LCB}^{a^*}>\text{UCB}^{a_0}$, sample from $a^*$ and $a^{**}$; otherwise return them. Then compute and track the any-time p-value $P^{(t)}$.
    \item Record $R^{(t)}=1\big[P^{(t)} > \alpha^{(t)}\big]$, and decide whether to stop based on $R^{(t)}$ or other stopping criteria.
\end{enumerate}

\section{Challenges of adaptive online testing}
\label{sec:difficulty}
Although the recent advancements can refine the procedure of adaptive online testing, they do not change the nature of the problem, especially its inherent difficulty.
Here, the difficulty refers to the minimum number of samples required by whatever smart algorithm to achieve high-confidence decisions.
For simplicity, assuming that each arm has unit variance. For uniform exploration, it is obvious that the difficulty lies exactly in how separable the set of arms are (in terms of $\mu_a$). 
Adaptive explorations may help us eliminate inferior arms faster, but the optimality of the eventual performance will depend more heavily on the gap between $a^{[1]}$ and $a^{[2]}$.
In general, the distribution patterns of the suboptimality gaps can induce different levels of difficulty for exploration, and the recently \citet{chen2017towards} reveals a (tight) lower bound to justify this phenomenon. 

In the theoretical literature, information-theoretical lower bound if often used to describe the amount of information needed to specify the answer related a particular problem instance. In the case of best-arm identification, the problem instance can be effectively described by the set: $\Ical=\big\{\Delta_{[j]}\big\}_{j=2}^k$, where $\Delta_{[j]}:=\mu_{a^{[1]}} - \mu_{a^{[j]}}$. 
\begin{lemma}[Difficulty of best-arm identification.]
\label{lemma:lower-bound}
Let $T_{\Dbb}(\Ical, \delta)$ be the stopping time for applying $\Dbb$ on problem $\Ical$ such that it achieves the $\delta$-correctness defined in (\ref{eqn:best-arm-id}). It holds that:
\begin{equation*}
    \inf_{\Dbb}T_{\Dbb}(\Ical, \delta) = \Ocal\Big(\sum_{j=2}^k\frac{1}{\Delta_{[j]}^2}\Big(\text{ln}\frac{1}{\delta} + H(\Ical) \Big) + \frac{\text{lnln}\Delta_{[2]}^{-1}}{\Delta_{[2]}^2} \text{polylog}(k,\delta^{-1})\Big),
\end{equation*}
where $H(\Ical)$ is an entropy term of the set $\Ical$.
\end{lemma}

Before we discuss this result, recall that adaptive exploration does not account for the asymmetry of the null and alternative hypothesis. \citet{yang2017framework} recently propose to give the control arm a little bit of edge by manually increasing its estimation by some $\epsilon>0$ at all time. Note that this procedure can change the gaps by at most $\epsilon$, so our following discussions are not affected by this modification.

There are three quantities essential to the lower bound: 1). the term $\sum_{j=2}^k 1/\Delta_{[j]}^2$ which measures the \emph{overall complexity} (total gap) of the problem instance; 2). the gap $1/\Delta_{[2]}^2$ between the best arm and the strongest contender; 3). the amount of variation (or divergence) in the gaps' distribution, as reflected by $H(\Ical)$. In general, the problem becomes more difficult when the gaps (especially $\Delta_{[2]}$) become smaller or more diverged. 

The above result leads to our discussion on the challenges of adaptive online testing. Intuitively, adaptive exploration will be more efficient as we increase the number of arms.
But now we understand how adding arms can also boost the complexity (either $\sum_{j=2}^k 1/\Delta_{[j]}^2$ or $1/\Delta_{[2]}^2$) or the entropy of the gaps. 
Consider a simple example with three arms and $\Delta_{[1]}=1, \Delta_{[2]}=1$. Suppose a new arm sits right on the middle of $a^{[1]}$ and $a^{[2]}$ and the gaps become: $\Delta_{[1]}=1/2, \Delta_{[2]}=1/2, \Delta_{[3]}=1$. It means adding this new arm makes the problem four times more difficult. Non-adaptive online testing might just be as efficient in this case. 

Not knowing the gaps in advance may not the the only factor that undermines adaptive online testing. Note that we previously assumed all arms have unit variance, which is unrealistic. Indeed, the variance associated with arm also plays a non-negligible role for the separability, and it actually acts as multiplicative factors on $\big\{1/\Delta_{[j]}^2\big\}_{j=2}^k$ \cite{kaufmann2016complexity}.
Further, the confidence level will also become a stochastic process (such as the $\alpha^{(t)}$ in Section \ref{sec:advances}) instead of the fixed $\delta$ we just discussed in Lemma \ref{lemma:lower-bound}.
The stochasticity in confidence level may further complicate the lower bound, and this scenario still requires new technical results to analyze. In light of these facts, conclusions cannot be made on how adaptive online testing is more effective than traditional A/B/n testing and why.

In conclusion, while recent advances from multiple domains have made adaptive online testing a practical and scientific alternative to traditional A/B/n testing, the inherent difficultly of the problem still poses challenges to fully embrace adaptive testing. Nevertheless, this technology has the potential to benefit multiple industries as more research and empirical work are underway.

\section{Future research directions}
We briefly discuss several promising directions that may contribute greatly to decision-making with adaptive online testing for IR.

\begin{itemize}[leftmargin=*]
    \item Extension to the fixed budget setting. The fixed budget setting may be more practical for certain IT productions such as advertising \cite{sen2017identifying}. While the best-arm identification algorithms can often switch between the fixed confidence and fixed budget setting \cite{gabillon2012best}, the focus of hypothesis testing may vary. Increasing the power of a experiment design, rather than controlling the false discovery rate, might be more demanding for fixed budget problems. More research is needed to incorporate the new objective.
    
    \item \emph{Thompson Sampling} (TS) for exploration. The Bayesian idea behind TS's exploration strategy is as powerful as the CB method \cite{agrawal2012analysis}, and \citet{russo2016simple} recently developed TS-driven best-arm identification algorithm. A distinctive advantage of TS is to include prior knowledge about the experimentation (e.g. offline testing results). Although hypothesis testing is strictly a frequentist's choice, there is possibility to seek collaboration especially when TS involves only in exploration.
    \item Involving \emph{contextual information}. Just like how contextual information serves multi-armed bandit, they can also be used to optimize adaptive online testing. For instance, when an online testing is designed to select the best banner size, the unknown rewards for each arm may be size-related. If so, an extra learning step can lead to better exploration. \citet{kato2021role} and \citet{chen2020contextual} has done some preliminary work in this direction, but this field remains largely unstudied.
    \item Real-world examination. Most of the methods in this short paper have only been tested via simulation or small-scale studies. Their efficacy in large-scale IT productions remains to be told.
    The real-time computation and updates required by adaptive online testing will certainly put a test on the infrastructure at the serving end. Business requirements may also complicate the design and analysis of adaptive online testing. Applied research with deployment solutions will be highly appreciated.
\end{itemize}

\bibliographystyle{ACM-Reference-Format}
\bibliography{references}

\end{document}